\documentclass[pra,aps,twocolumn,showpacs,superscriptaddress,nofootinbib]{revtex4}

\input{Qcircuit}
\input epsf

\begin{document}

\title{Noise thresholds for optical quantum computers}

\author{Christopher M. Dawson}
\affiliation{School of Physical Sciences,
The University of Queensland, Queensland 4072, Australia}

\author{Henry L. Haselgrove}
\affiliation{School of Physical Sciences, The University of
Queensland, Queensland 4072, Australia}
\affiliation{Information
Sciences Laboratory, Defence Science and Technology Organisation,
Edinburgh 5111 Australia}
\author{Michael A. Nielsen}
\affiliation{School of Physical Sciences,
The University of Queensland, Queensland 4072, Australia}

\date{\today}

\begin{abstract}
  In this paper we numerically investigate the fault-tolerant
  threshold for optical cluster-state quantum computing.  We allow
  both photon loss noise and depolarizing noise (as a general proxy
  for all local noise), and obtain a \emph{threshold region} of
  allowed pairs of values for the two types of noise.  Roughly
  speaking, our results show that scalable optical quantum computing
  is possible for photon loss probabilities $< 3 \times 10^{-3}$, and
  for depolarization probabilities $< 10^{-4}$.
\end{abstract}


\maketitle

Optical systems are promising candidates for quantum computation, due
to their long decoherence times, accurate single-qubit gates, and
relatively efficient readout.  A scheme for optical quantum computing
has been suggested by Knill, Laflamme and Milburn
(KLM)~\cite{Knill01a}, and the basic elements of that scheme
experimentally
demonstrated~\cite{Pittman03a,OBrien03a,Sanaka03a,Zhao04a}.
Unfortunately, KLM requires tens of thousands of optical elements to
achieve a \emph{single} entangling gate operating with high
probability.  A recent proposal~\cite{Nielsen04b}
(c.f.~\cite{Yoran03a}) combines elements of KLM with the cluster-state
model of quantum computation~\cite{Raussendorf01a} to reduce the
complexity by many orders of magnitude.  This scheme has been
simplified~\cite{Browne05a} to require only tens of optical elements
per logical gate.  A recent experiment~\cite{Gasparoni04a}
demonstrated simple optical cluster state computing.

Our paper investigates the effect of noise on the optical cluster
state proposals.  In the standard quantum circuit model, the noise
threshold theorem shows that provided the amount of noise per
elementary operation is below the threshold, scalable quantum
computation is possible.  Unfortunately, this result does not apply
directly to optical cluster states, as the cluster state model is
fundamentally different from the circuit model.
However,~\cite{Nielsen05a,Raussendorf03b} (c.f.~\cite{Tame05a})
established the \emph{existence} of a threshold for clusters, without
obtaining a value, while~\cite{Aliferis05a} argued that in a certain
noise model the cluster threshold is no more than an order of
magnitude worse than the circuit threshold.  This latter work is not
directly relevant to optical clusters, since it uses deterministic
entangling gates, and does not include any process analogous to photon
loss.

We use numerical simulations to find the noise threshold.  Our
analysis is tailored to the dominant sources of noise in optical
quantum computing, including the nondeterminism of the optical
entangling gates, photon loss, and depolarizing noise. We therefore
obtain a threshold \emph{region} of noise parameters for which
scalable optical quantum computing is possible.  Our protocol is
complex, and we omit some details; full details will appear
in~\cite{Dawson05a}.

A prior work~\cite{Varnava05a} has calculated a threshold for optical
quantum computation when the only source of noise is photon loss.  In
real experiments other noise sources such as dephasing are also
present, and protocols such as~\cite{Varnava05a} will actually amplify
the effects of such noise at the encoded level.  By contrast, our
protocol protects against both photon loss and depolarizing noise, and
by standard fault-tolerance results thus automatically protects
against arbitrary local noise, including dephasing (in any basis),
amplitude damping, etc.

Introductions to cluster-state computation may be found
in~\cite{Nielsen05b,Raussendorf03a}, and we assume familiarity with
the model.  An important element in the model are the Pauli
``byproduct'' operators, known functions of the measurement results,
which are used to correct the state when the computation concludes.
We call the tensor product of these byproduct operators the
\emph{Pauli frame}, and it is updated after measurements of cluster
qubits according to a set of \emph{propagation rules}, described,
e.g., in~\cite{Nielsen05b}.

Our approach to optical cluster-state computation is based most
closely on~\cite{Browne05a}. We use the polarization of a single
photon to encode a single qubit, and build clusters up using
\emph{fusion gates} (``type I fusion gates'' in~\cite{Browne05a}),
which, when applied to two cluster qubits either (a) fuse the qubits
into a single cluster qubit, which occurs with probability
$\frac{1}{2}$; or (b) measure both qubits in the computational basis,
also with probability $\frac{1}{2}$.

\textbf{Available physical resources and noise:} Our resources are:
(1) a source of polarization entangled Bell pairs; (2) single-qubit
gates, effected using linear optics, as in KLM; (3) efficient
polarization-discriminating photon counters capable of distinguishing
$0, 1$ and $2$ photons; (4) fusion gates, built from beamsplitters and
photon counters; and (5) quantum memory gates. We assume all these
elements take the same amount of time, and describe our circuit as a
sequence of such time steps.

Our noise model includes a parameter $\gamma$ representing the
probability per qubit per time step of photon loss.  We assume this
probability is independent of the state of the qubit, and that photon
loss occurs after Bell-state preparation, and before memory,
single-qubit and fusion gates; for two-qubit operations we assume
photon loss occurs independently for both qubits.  Our noise model
also includes a \emph{depolarizing parameter} $\epsilon$.
Depolarizing noise affects physical operations as follows: (1) after
Bell-state preparation or before a fusion gate the two qubits are
collectively depolarized, i.e., with probability $1-\epsilon$ nothing
happens, while with probabilities $\epsilon/15$ we apply each of the
$15$ non-identity Pauli operators $IX, XX$ etc; and (2) before memory,
single-qubit and measurement gates the qubit is depolarized with
parameter $\epsilon$.



Additional noise sources that may effect real implementations include
dark counts and dephasing. However, the fault-tolerant protocol we
implement automatically protects against such noise sources, and we
believe the threshold results will not qualitatively change.

\textbf{Method of simulation:} We use the stabilizer formalism to
simulate Clifford group operations, which are sufficient to simulate
error-correction and depolarization. In our simulations, rather than
working with the state directly, we merely keep track of the
\emph{errors} in the state when compared with an ideal reference
state.  We keep track of two types of errors: the physical error in
the state of the cluster, which is represented as a tensor product of
Pauli operators, and errors in the Pauli frame due to erroneous
measurement results, which are again a tensor product of Pauli
operators.  Note that physical errors may propagate to become Pauli
frame errors when qubits with physical noise are measured, giving rise
to incorrect measurement results.  The rules for propagating both
types of errors may be computed following, e.g.,~\cite{Nielsen05b};
see~\cite{Dawson05a}.

These methods suffice to describe error-correction and Pauli-type
noise, but not photon loss and fusion gate failures.  We can use
postselection and repetition to effectively eliminate photon loss and
fusion gate failures, whenever those failures do not directly affect
the encoded data.  However, when they do affect the data, another
approach must be taken.  Suppose when fusion gate failure occurs the
experimenter: (1) Randomizes the local Pauli frame of the data qubit;
(2) notes the location at which the failure occurred, for use in
decoding; and (3) carries out the rules for propagating the Pauli
frame, as though the fusion gate had succeeded, and the bond was
created.  The rules for propagating Pauli frame errors can be used to
show that once the experimenter has randomized the Pauli frame, it
does not make any physical difference whether the fusion gate failed
or not, and so we can treat it as though it succeeded.  The remaining
errors are Pauli-type errors, and so can be simulated in the standard
way.  The details (and a discussion of photon loss, which is dealt
with similarly) appear in~\cite{Dawson05a}.

\textbf{Broad picture of fault-tolerant protocol:} The protocol is
split into two parts: (1) a cluster-based simulation of a variant of
Steane's protocol~\cite{Steane03a}; (2) a deterministic gate-based
protocol, again based on~\cite{Steane03a}.  The cluster threshold is
obtained by concatenating the results from a single level of the
cluster protocol with multiple levels of the deterministic protocol.
The idea is to take a quantum circuit, build up a fault-tolerant
simulation through multiple levels of concatenation in the circuit
model, and then replace the bottom level by a clusterized simulation
of a noisy deterministic gate.

\textbf{Microclusters and parallel fusion:} Our protocol is based on
\emph{microclusters}, star-shaped clusters with a central \emph{root
  node}, and attached \emph{leaf nodes}.  Such microclusters can be
created using repeated fusion of Bell pairs.  By attempting
preparation of a large number of microclusters in parallel and
postselecting on successful attempts we can build a $k$-leaf
microcluster in $O(\log(k))$ timesteps, and consuming $O(k^2)$ Bell
pairs, with probability of success arbitrarily close to one.

Microclusters can be used to ensure that larger clusters always have
multiple leaf nodes. This can be used to enhance the probability of
fusing two clusters, by attempting simultaneous fusion gates between
adjacent leaf nodes of the two clusters.  With a probability that goes
rapidly to one as the number of leaves increases, at least one of
these fusion gates succeeds, fusing the two clusters together. We call
this process of using leaves to fuse the two clusters with high
probability \emph{parallel fusion}.

\textbf{State at the start of a round:} The ideal noise-free state at
the start of any round is the encoded state of the data (e.g., in the
$7$-qubit code), but with a number of leaves attached to each code
qubit, which are used later for parallel fusion.


At the beginning of the entire trial, we assume the input is a
noiseless state of this form. We justify this assumption on the
grounds that the initial state does not actually matter, since our
goal is to estimate the rate \emph{per round} at which crashes occur
in the encoded data.  Following~\cite{Steane03a}, we perform warm-up
rounds of error correction before gathering data on this crash rate,
so as to avoid transient effects due to the choice of initial state.

\textbf{Ancilla creation:} Each round of error correction involves the
creation of multiple verified ancilla states, which are used to
extract syndrome bits.  We illustrate this for the $7$-qubit code, but
the procedure generalizes to other codes.  We create the ancilla using
the cluster:
\begin{equation}
\epsfxsize=6cm \epsfbox{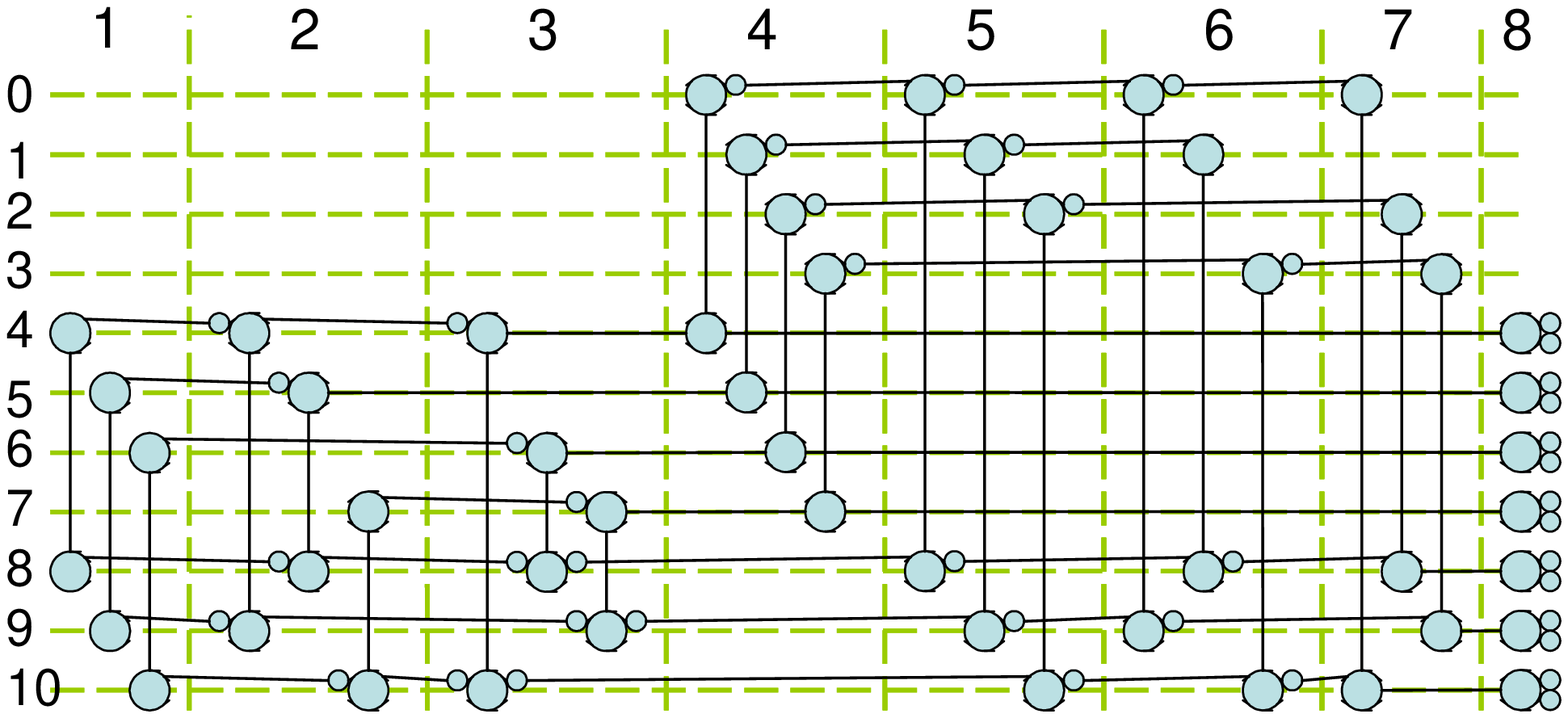}
\end{equation}
This is a clusterization of the ancilla creation circuit
in~\cite{Steane03a}, with the final column of the cluster
corresponding to an encoded $|+\rangle$.  We abridge our notation so
that touching circles represent connected cluster qubits.  The cluster
is created by first creating an array of microclusters; the large
circles represent root nodes, while the smaller circles represent
leaves; note that many of the leaves are consumed during preparation
by fusion and parallel fusion, and are not shown.  We then use fusion
and parallel fusion to create the bonds; details appear
in~\cite{Dawson05a}.  We conclude by measuring all qubits in the $X$
basis, except the leaves in column $8$.  To verify the ancilla, we
postselect on the measurement results of the terminating qubits in
rows $0, 1, 2, 3$ all being $0$.  The resulting state is an encoded
$|+\rangle$, with each qubit having a number of leaves attached for
the purpose of parallel fusion.


\textbf{The telecorrector:} To extract error syndromes we interact the
data and ancillas using a special cluster state called a
\emph{telecorrector}.  Telecorrector-based syndrome extraction is a
variant of Steane's approach (c.f. also the related protocol
in~\cite{Knill05a}), and can be thought of as a clusterized version of
the circuit:
\begin{equation} \label{eq:steane_extraction}
\epsfxsize=5.5cm \epsfbox{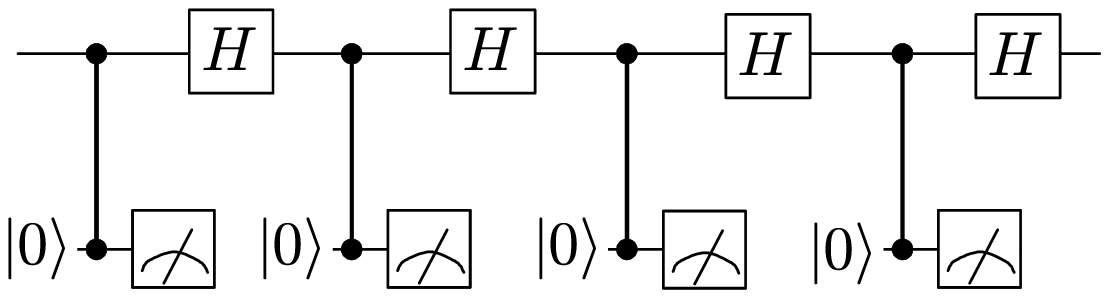}
\end{equation}
where operations are being performed on encoded qubits, $|0\rangle$ is
fault-tolerant ancilla creation, and the measurement is a transversal
$X$ basis measurement.

Telecorrector creation begins with the creation of multiple copies of
the following state, one copy for each qubit in the code being used:
\begin{equation} \label{eq:telemodule}
\epsfxsize=4.5cm \epsfbox{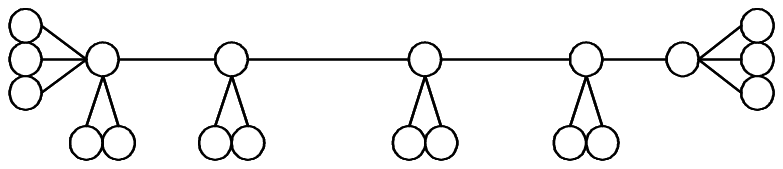}
\end{equation}
This state can be created in the obvious way using postselected
microcluster fusion.  The leaves on the left-hand end will
eventually be used to attach to a single qubit of the encoded data
using parallel fusion.  The leaves and root node on the right-hand
end will contain the output of this round of error-correction, and
become the input to the next round of error-correction.  The
remaining leaves will be used to fuse to ancilla states.

Simultaneous with the creation of the state~(\ref{eq:telemodule}), we
create four verified ancilla states, and fuse the ancillas with the
leaves on the state~(\ref{eq:telemodule}) to create the state
(illustrated as though for a three-qubit code):
\begin{equation} \label{eq:telecorrector}
\hspace{-1ex} \epsfxsize=7.5cm \epsfbox{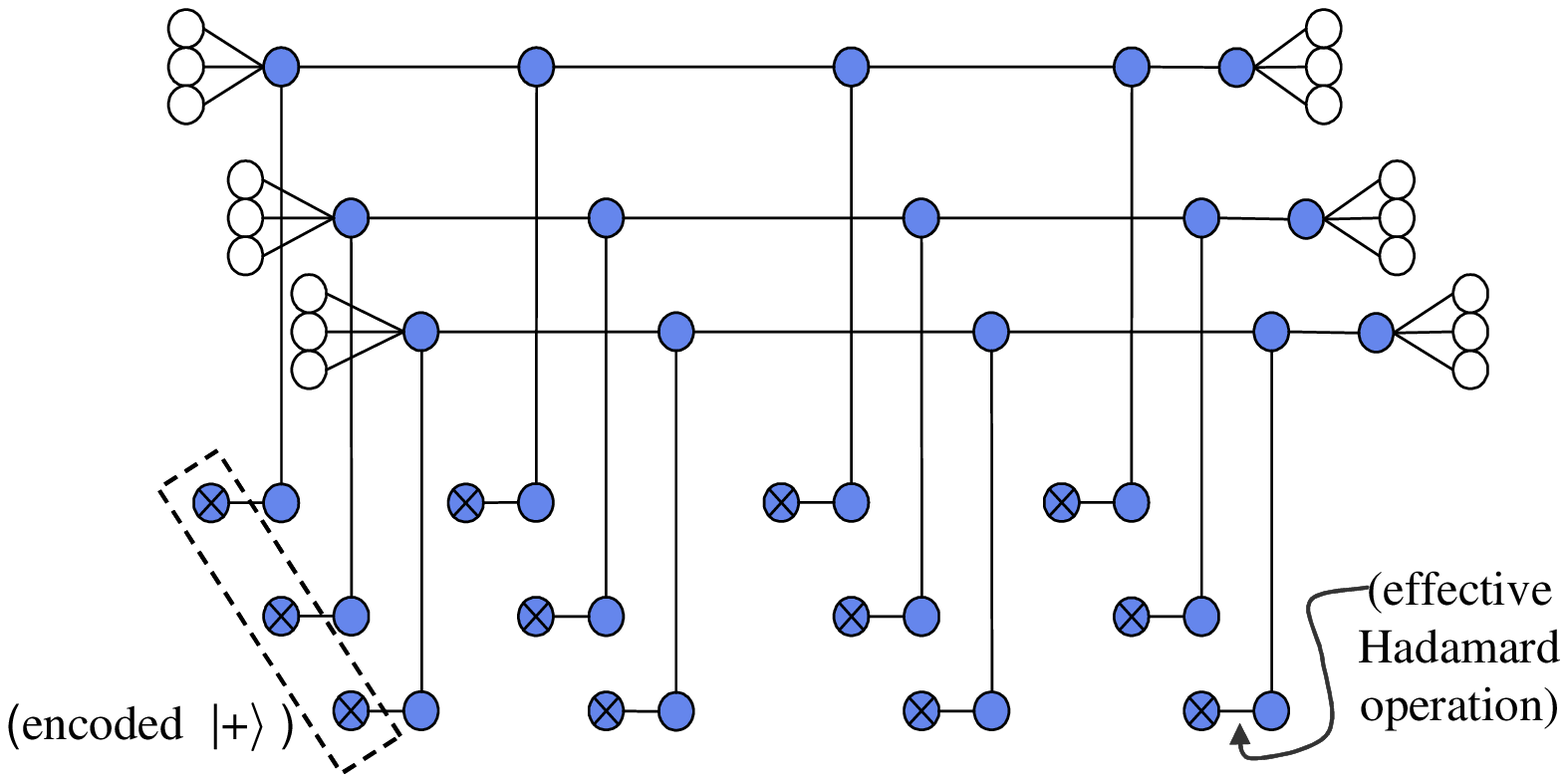}
\end{equation}
Next we measure all the shaded qubits in the $X$ basis, leaving only
the leftmost and rightmost leaves, for later use in attaching the
data, and future rounds of error correction.  Applying the propagation
rules for the Pauli frame, it can be shown that the measurement
outcomes from the shaded qubits completely determine whether the
repeated syndrome measurements will agree or not, \emph{before the
  state has interacted with the data}!  We have verified this fact
both numerically and analytically.

We take advantage of this remarkable fact by postselecting on
measurement outcomes that ensure this \emph{preagreeing syndrome}
property.  We call the resulting state the \emph{telecorrector}.  The
preagreeing syndrome property enables syndrome extraction to be
performed more efficiently (thus improving the threshold) than in
Steane's protocol, which extracts many syndromes to ensure that some
large subset agree.

Once prepared, we use parallel fusion to attach the telecorrector to
the data, and then $X$ basis measurements to complete this part of the
computation.  Standard propagation rules are used to update the Pauli
frame, and to determine the syndrome extracted from this procedure.


\textbf{Decoding:} We use a technique for syndrome decoding which
takes advantage of the experimenter's knowledge of the locations of
photon loss and fusion gate failures.  In particular, we use the fact
(see p.~467 of~\cite{Nielsen00a}) that a code correcting $t$ unlocated
errors is able to correct $2t$ located errors.  Our technique is a
maximum likelihood procedure for decoding arbitrary combinations of
located and unlocated errors.  All the codes we use are CSS codes with
the property that decoding of the $X$ and $Z$ errors can be performed
separately using an identical procedure.  The $X$-decoding procedure
(for example) has the following inputs: the measured $X$-error
syndrome, obtained from the vector of total errors of the ancilla
measurement outcomes; and a list of locations (qubit indices within
the code block) at which located errors have occurred during the
round. The outputs of the decoding routine are: a list of locations
where $X$ flips should be made in order to correct the data; and a
flag signalling a \emph{located crash}. The located crash flag is set
to ``true'' when different patterns of $X$ errors are found to have
equal maximum likelihood, but differ from each other by a logical $X$
operation. The located crash flag is used to improve decoding at the
next level of concatenation, by identifying encoded blocks which are
known to have experienced an error.

\textbf{Results of the optical cluster simulation:} Our aim in
simulating cluster-based error correction is to estimate the function
which maps the input noise parameters $(\epsilon, \gamma)$ to the
logical error rates, or {\em crash} rates, defined below. Below we
describe simulations which estaimate a similar function for a
deterministic circuit-based protocol, and then combine the results to
give the threshold curve for cluster-state optical quantum computing.

At the end of a round of simulated cluster-based error correction, we
say that the round has caused a {\em located crash} whenever either
the $X$ or $Z$ decoding steps have reported a located crash.
We define an {\em unlocated crash} as follows. We take the pattern of
Pauli errors on the root nodes of the data, and consider the result of
a perfect (noise-free) round of correction.  If perfect correction
would result in a pattern on Pauli errors corresponding to a
non-identity encoded Pauli operation, then we say the data has
experienced an unlocated crash.


We performed simulations based on the Golay 23-qubit and Steane
7-qubit codes.  For each simulation, we chose a number of settings for
the noise parameters $(\epsilon,\gamma)$, and for each we ran a
many-trial monte carlo simulation. Each trial consisted of two
successive rounds of error correction, and the outcome of the trial
was determined by whether the second of the two rounds caused a crash.
The purpose of the first ``warm-up'' round is to reduce transient
effects due to our choice of (noise free) initial conditions.
Including more than one warm-up round did not make a statistically
significant change to the results.

We tally the outcomes as follows. For all the trials for which the
first round does not cause a crash, we count: (1) the number $N_U$ for
which the second round causes an unlocated crash but not a located
crash, (2) the number $N_L$ for which the second round causes a
located crash, and (3) the number $N_N$ for which no crashes occur.
The unlocated and located crash rates $E$ and $\Gamma$ are estimated
as $E=\frac{N_U}{N_U+N_N}$ and $\Gamma= \frac{N_L}{N_U+N_N+N_L}$.  We
calculate these crash rates for a variety of input noise parameters,
and use weighted least-squares fitting to fit polynomials
$E(\epsilon,\gamma)$ and $\Gamma(\epsilon,\gamma)$ representing the
general behaviour of the crash rates; these polynomials were in good
agreement with the qualitative theory of fault-tolerance.

\textbf{Concatenation:} Under $k$ layers of concatenation, our
protocol is effectively equivalent to doing $k-1$ concatenated levels
of an ordinary deterministic gate-based fault-tolerance protocol, and
then replacing the elements at the lowest level by cluster-based
equivalents with just a single level of encoding.  To understand the
behaviour of the concatenated protocol, we therefore also did
simulations of a deterministic fault-tolerant protocol.  These
simulations followed~\cite{Steane03a}, but incorporated a
deterministic version of telecorrection, and the separate treatment of
unlocated and located errors.  The appropriate noise model has two
noise parameters, $(p,q)$, representing, respectively, the rate of
located and unlocated Pauli errors, corresponding to located and
unlocated crashes at the next lowest level of concatenation.  The
results of our simulations suggest that these crashes may be
accurately modelled as independent $X$ and $Z$ errors, with $Y$ errors
suppressed, and so this is the noise model we adopt.  We use least
squares fitting to estimate polynomials $P(p,q)$ and $Q(p,q)$, where
$P$ and $Q$ are the rates for located and unlocated crashes at a
single level of encoding in the cluster-based simulations.

\begin{figure}[t]
\begin{center}
\epsfxsize=6cm \epsfbox{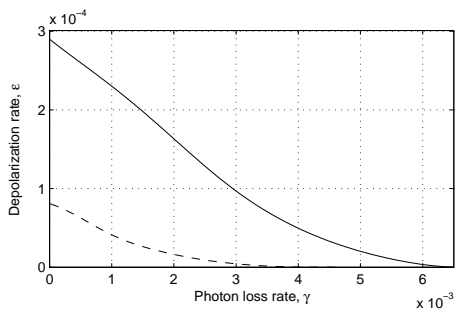}
\end{center}
\caption{Threshold region for optical clusters using the
  23-qubit Golay code (solid) and 7-qubit Steane code (dashed).
}\label{fig:oclus23}
\end{figure}

\textbf{Results and conclusion:} Define maps $f: (\epsilon,\gamma)
\rightarrow (E,\Gamma)$ and $g : (p,q) \rightarrow (P,Q)$.  Then the
located and unlocated crash rates after $k$ levels of concatenation
may be estimated by computing $(g^{(k-1)} \circ f)(\epsilon,\gamma)$.
Provided this tends to $(0,0)$ as $k \rightarrow \infty$ we are inside
the threshold region.  Fig.~\ref{fig:oclus23} illustrates the
threshold region, and shows that the 23-qubit code gives a
considerably better threshold than the 7-qubit code; resource usage
will be discussed in~\cite{Dawson05a}.  Both codes give thresholds
worse than the best known circuit thresholds~\cite{Knill05a}, but the
results are encouraging given the non-deterministic nature of the
optical entangling operations.

\acknowledgments

Thanks to Dan Browne, Ike Chuang, Jen Dodd, Steve Flammia, and Tim
Ralph for helpful discussions, and to Bryan Eastin and Steve Flammia
for \emph{Qcircuit}.


\end{document}